\documentclass[onecolumn,showpacs]{revtex4}

\usepackage{psfig}
\usepackage{epsfig}
\usepackage{graphicx}
\usepackage{dcolumn}
\usepackage{amsmath}

\makeatletter
\def\btt#1{\texttt{\@backslashchar#1}}%
\DeclareRobustCommand\bblash{\btt{\@backslashchar}}%
\makeatother

\begin{document}

\title{Tachyon Monopole}

\author{Xin-zhou Li}\email{kychz@shnu.edu.cn}
\author{Dao-jun Liu}

\affiliation{Shanghai United Center for Astrophysics(SUCA), Shanghai
Normal University, 100 Guilin Road, Shanghai 200234,China
}%

\date{\today}

\begin{abstract}
The property and gravitational field of global monopole of tachyon
are investigated in a four dimensional static space-time.  We give
an exact solution of the tachyon field in the flat space-time
background. Using the linearized approximation of gravity, we get
the approximate solution of the metric. We also solve analytically
the coupled Einstein and tachyon field equations which is beyond
the linearized approximation to determine the gravitational
properties of the monopole solution. We find that the metric of
tachyon monopole represents an asymptotically  AdS space-time with
a small effective mass at the origin. We show that this relatively
tiny mass is actually negative, as it is in the case of ordinary
scalar field.
\end{abstract}

\pacs{11.27.+d, 98.80.Cq, 11.25.-w}

\maketitle

\vspace{0.5cm}\noindent \textbf{1. Introduction} \vspace{0.5cm}

Recently, pioneered by Sen \cite{Sen1}, the study of non-BPS
objects such as non-BPS branes, brane-antibrane configurations or
space-like branes \cite{Strominger} has been attracting physical
interests in string theory. Sen showed that classical decay of
unstable D-brane in string theories produces pressureless gas with
non-zero energy density. The basic idea is that the usual open
string vacuum is unstable but there exists a stable vacuum with
zero energy density. There is evidence that this state is
associated with the condensation of electric flux tubes of closed
string \cite{Sen2}. These flux tubes described successfully using
an effective Born-Infeld action \cite{Sen3}.  The tachyon rolling
towards its minimum at infinity as a dark matter candidate was
also proposed by Sen \cite{Sen2}. Sen have also studied the
Dirac-Born-Infeld Action on the Tachyon Kink and
Vortex\cite{sen4}. Gibbons took into account coupling to
gravitational field by adding an Einstein-Hilbert term to the
effective action of the tachyon on a brane, and  initiated a study
of "tachyon cosmology" \cite{Gibbons}. Several authors have
investigated the process of rolling of the tachyon in the
cosmological background \cite{Li4, Kofman}.

Various topological defects such as domain wall, string and
monopole could be formed by the symmetry-breaking  phase
transitions in the early universe and their existence has
important implications in cosmology\cite{Kibble, vilenkin}. The
symmetry breaking model of ordinary scalar field can be
prototypically written as

\begin{equation}
\emph{L}
=\frac{1}{2}\partial_{\mu}\phi^{a}\partial^{\mu}\phi^{a}-V(f)
\end{equation}

\noindent where $\phi_{a}$ is a set of scalar fields, $a=1, ...,
N$, $f=(\phi^{a}\phi^{a})^{\frac{1}{2}}$. The model has O(N)
symmetry and admits domain wall, string and monopole solutions for
$N=1, 2$ and $3$, respectively \cite{Chen}.  Usually, the
potential $V(f)$ has a minimum at a finite non-zero value of $f$.
On the other hand, in Ref. \cite{Cho1, Cho2}, Cho and Vilenkin
investigated the defects in models where $V(f)$ has a local
maximum at $f=0$ but no minima; instead, it monotonically decrease
to zero at $f\rightarrow \infty$. And they called this kind of
defects "vacuumless defects".

Global monopole, which has a global O(3) symmetry spontaneously
broken to U(1), is one of the most interesting above mentioned
defects. The property of the global monopole in curved space-time,
or equivalently, its gravitational effects, was firstly studied by
Barriola and Vilenkin \cite{barriola}. When one considers the
gravity, the linearly divergent mass of global monopole has an
effect analogous to that of a deficit solid angle plus that of a
tiny mass at the origin. Harari and Loust\`{o} \cite{harari}, and
Shi and Li \cite{li1} have shown that this small gravitational
potential is actually repulsive. A new class of cold stars,
addressed as D-stars(defect stars) have been proposed by Li
et.al.\cite{li2, li3}. One of the most important features of such
stars, comparing to Q-stars, is that the theory has monopole
solutions when the matter field is absent, which makes the D-stars
behave very differently from the Q-stars. Furthermore, if the
tachyon field $T$ rolls down from the maximum of its potential and
the universe does not inflate, the quantum fluctuations produced
various topological defects during spontaneous symmetry breaking. It
is therefore of importance to investigate the property and the
gravity of the topological defects of tachyon, such as vortex
\cite{liu}, kink \cite{ckim} and monopole, in the static space-time.
In the next section, equations of tachyon monopole are presented. In
sect. 3, we find an analytical solution of tachyon monopole in the
flat space-time background and discuss its property. We show that
because the tachyon field rolls down to its ground without
oscillating about it, the tachyon monopole becomes very diffuse,
just as the "vacuumless monopole" studied in Ref. \cite{Cho1}. In
sect. 4, the gravitational field of tachyon monopole are
investigated by using linearized gravity approximation and
considering the coupled Einstein and tachyon field equation which is
beyond linearized approximation, respectively. We will find that the
metric of tachyon monopole represents an asymptotically AdS
space-time with a small effective mass at the origin which is
actually negative. If tachyon monopole existed, they would share
with other topological defects, such as domain walls and strings,
curious and rather unconventional gravitational effect. Finally, we
give some brief discussions.

\vspace{0.5cm}\noindent \textbf{2. Equations of motion}
\vspace{0.5cm}

A general static, spherically-symmetric metric can be represented
as

\begin{equation}\label{metric}
ds^2=B(r)dt^2-A(r)dr^2-r^2d\Omega^2
\end{equation}

\noindent where $d\Omega^2$ is the metric on a unit 2-sphere. And
the Lagrangian density of rolling tachyon with potential
$V(\phi)$, which couples to the Einstein gravity, can be written
as the following Born-Infeld form:

\begin{eqnarray}\label{LD:2}
L&=&L_R+L_T\nonumber\\
&=&\sqrt{-g}\bigg[\frac{R}{2\kappa}-V(|T|)\sqrt{1-g^{\mu\nu}\partial_{\mu}T^{a}\partial_{\nu}T^{a}}\bigg]
\end{eqnarray}

\noindent where $T^{a}$ is a triplet of  tachyon fields, $a=1,2,3$
and $g_{\mu\nu}$ is the metric. The field configuration describing
a monopole is given by

\begin{equation}\label{ansatz:3}
T^a=f(r)\frac{x^a}{r}
\end{equation}

\noindent where $x^ax^a=r^2$. From the Lagrangian density
(\ref{LD:2}), metric (\ref{metric}) and the above configuration of
the field(\ref{ansatz:3}), we can obtain the following
Euler-Lagrange equation:

\begin{equation}\label{ELequation1}
\frac{1}{V}\frac{dV}{df}+\frac{2f}{r^2}=\frac{1}{A}\bigg[f''+(\frac{B'}{2B}-\frac{A'}{2A}+\frac{2}{r})f'-f'\frac{\frac{f'f''}{A}-\frac{f'^2A'}{2A^2}+\frac{2f}{r^2}(f'-\frac{f}{r})}{1+\frac{f'^2}{A}+\frac{2f^2}{r^2}}\bigg]
\end{equation}

\noindent where the prime denotes the derivative with respect to
$r$. The Einstein equations read:

\begin{equation}\label{EinsteinEq1}
\frac{1}{r^2}-\frac{1}{A}(\frac{1}{r^2}+\frac{1}{r}\frac{A'}{A})=\kappa
T^0_0
\end{equation}

\begin{equation}\label{EinsteinEq2}
\frac{1}{r^2}-\frac{1}{A}(\frac{1}{r^2}+\frac{1}{r}\frac{B'}{B})=\kappa
T^1_1
\end{equation}

\noindent where the energy-momentum tensor $T^{\mu}_{\nu}$ of the
system are given by

\begin{equation}\label{t00}
T^0_0=V(f)\sqrt{1+\frac{f'^2}{A}+\frac{2f^2}{r^2}}
\end{equation}

\begin{equation}\label{t11}
T^1_1=\frac{V(f)(1+\frac{2f^2}{r^2})}{\sqrt{1+\frac{f'^2}{A}+\frac{2f^2}{r^2}}}
\end{equation}

\begin{equation}\label{t22}
T^2_2=T^3_3=\frac{V(f)(1+\frac{f'^2}{A}+\frac{f^2}{r^2})}{\sqrt{1+\frac{f'^2}{A}+\frac{2f^2}{r^2}}}
\end{equation}

\noindent and the rest are zero. Obviously, the configuration of
the system depends on the tachyon potential $V(T)$. However, the
explicit form of the potential is not clear now. According to Sen
\cite{Sen2}, the potential should have an unstable maximum at
$T=0$ and decay exponentially to zero when $T\rightarrow\infty$.
There are lots of functional forms that satisfy the above two
requirements. In the following sections, we choose the tachyon
potential as follows:

\begin{equation}\label{V1}
V(f)=M^4(1+3\lambda f^4)^{1/6}\exp(-\lambda f^4)
\end{equation}

\noindent where $M$ and $\lambda$ are two constants and both are
greater than zero. Obviously, this potential dacays faster than
the exponential one for large $f$, but qualitatively, it has the
similar property as the exponential potential.

\vspace{0.5cm} \noindent \textbf{3. The analytical solution of
tachyon monopole in flat space-time background} \vspace{0.5cm}

In flat space-time background, the Euler-Lagrange equation
(\ref{ELequation1}) can be reduced to the following equation:

\begin{equation}\label{ELequation2}
\frac{1}{V}\frac{dV}{df}+\frac{2f}{r^2}=f''+\frac{2f'}{r}-f'\frac{f'f''+\frac{2f}{r^2}(f'-\frac{f}{r})}{1+f'^2+\frac{2f^2}{r^2}},
\end{equation}

\noindent and the energy density of the system is written as

\begin{equation}\label{energyD}
T^0_0=V(f)\sqrt{1+f'^2+\frac{2f^2}{r^2}}.
\end{equation}

For the potential (\ref{V1}), the equation (\ref{ELequation2}) has
a simple exact solution

\begin{equation}\label{f1}
f(r)=\lambda^{-\frac{1}{4}}\bigg(\frac{r}{\delta}\bigg)^{-1},
\end{equation}

\noindent  where $\delta = \lambda^{-1/4}$ is the size of the
monopole core, and the corresponding energy density
(\ref{energyD}) can be rewritten as

\begin{equation}
T^0_0=M^4\bigg[1+3\bigg(\frac{\delta}{r}\bigg)^4\bigg]^{2/3}\exp\bigg[-\bigg(\frac{\delta}{r}\bigg)^4\bigg].
\end{equation}

The total energy of the tachyon monopole is

\begin{equation}\label{mu17}
\mu(R)= 4\pi \int^{R}_0 T^0_0 r^2dr \sim 4\pi
M^4\bigg(\frac{R^3}{3}-\frac{1}{\lambda R}\bigg),
\end{equation}

\noindent where the cutoff radius $R$ is set by the distance to
the nearest  anti-monopole. We find easily from Eq.(\ref{mu17})
that tachyon monopoles are very diffuse objects with most of the
energy distributed at large distances from the monopole core, and
we expect their space-time to be substantially distinct from the
ordinary case. They are much more diffuse than ordinary global
monopoles which have $\mu(R)\propto R$, so that the gravitational
effect of this configuration is equivalent to that of a deficit
solid angle in the metric with a negative mass at the origin.
Furthermore, they are similar to the vacuumless monopoles which
have $\mu(R)\propto R(R/\delta)^{4/(n+2)}$, so that most of the
energy is also diffused at large distance \cite{Cho1}. By analogy
with vacuumless monopoles and ordinary monopoles, the typical
distance between the monopoles at time $t$ is $R\sim t$
\cite{Cho1}, apart from the close monopole-antimonopole pairs
which are about to annihilate. Like ordinary monopoles, one of the
cosmological bound to tachyon monopole is that it should not
overclose the universe. This means that the energy density in
monopoles $\rho_m$ today should be less than the critical density
of the present universe. The ratio of monopole energy density to
critical density $\Omega_m$ at time $t$ is

\begin{equation}
\Omega_m(t)=\frac{\rho_m}{\rho_c(t)}\sim \frac{32\pi M^4 t^2}{3
M_p^2}
\end{equation}

\noindent where $M_p$ is the Planck mass. Therefore, the parameter
$M$ should satisfies the condition $M\lesssim 10^{-3}\mbox{eV}$ in
order to avoid conflicting present cosmological observations.

\vspace{0.5cm}\noindent \textbf{4. The gravitational field of
tachyon monopole}\vspace{0.5cm}

Firstly, the space-time of tachyon monopole will be studied using
the linearized gravity approximation in order to show the main
features  of its metric in a simple manner. And then, we will
solve analytically the coupled Einstein and tachyon field
equations which is beyond the linearized approximation to
rigorously determine the gravitational properties of the monopole
solution.

 \vspace{0.5cm}\noindent
\textbf{4.1 Linearized approximation}\vspace{0.5cm}

In this subsection, we study the space-time of global  tachyon
monopole using the linearized gravity approximation. We shall
first consider the Newtonian approximation. The Newtonian
potential $\Phi$ can be found for the equation

\begin{equation}\label{NewtonEq19}
\nabla^2 \Phi=\frac{\kappa}{2}(T^0_0-T^i_i).
\end{equation}

\noindent For the tachyon monopole, $f(r)$ is given by
Eq.(\ref{f1}) and

\begin{equation}
T^0_0-T^i_i \simeq -2M^4
\end{equation}

\noindent at $r>>\delta$. Then the solution of Eq.
(\ref{NewtonEq19}) is

\begin{equation}
\Phi(r) \simeq -\frac{4\pi M^4}{3\lambda M_p^2f^2}.
\end{equation}

\noindent The linearized approximation applies as long as
$|\Phi(r)|<<1$, which is equivalent to
$f>>\sqrt{\frac{4\pi}{3\lambda}}\frac{M^2}{M_p}$. Therefore, we
should require the parameters $\lambda$ and $M$ satisfy

\begin{equation}\label{21}
\lambda^{-1/4}<<R<<\sqrt{\frac{3}{4\pi}}\frac{M_p}{M^2}
\end{equation}

\noindent where $R$ is the distance to the nearest anti-monopole.
Next, we express the metric functions $A(r)$ and $B(r)$ as

\begin{equation}
A(r)=1+\alpha(r), \hspace{1cm} B(r)=1+\beta(r).
\end{equation}

\noindent Linearizing in $\alpha(r)$ and $\beta(r)$, and using the
flat space expression (\ref{f1}) for $f(r)$, the Einstein
equations (\ref{EinsteinEq1}) and (\ref{EinsteinEq2}) can be
written as follows

\begin{equation}\label{LEE1}
\frac{\alpha'}{r}+\frac{\beta'}{r}=\kappa M^4\bigg(\frac{\delta}{
r}\bigg)^4\bigg[1+3\bigg(\frac{\delta}{r}\bigg)^4\bigg]^{-1/3}\exp\bigg[-\bigg(\frac{\delta}{
r}\bigg)^4\bigg]
\end{equation}

\noindent and

\begin{equation}\label{LEE2}
\beta''+\frac{2\beta'}{r}=-\kappa
M^4\bigg[2+3\bigg(\frac{\delta}{r}\bigg)^4\bigg]\bigg[1+3\bigg(\frac{\delta}{r}\bigg)^4\bigg]^{-1/3}\exp\bigg[-\bigg(\frac{\delta}{
r}\bigg)^4\bigg].
\end{equation}

\noindent The solution of external metric is easily found

\begin{equation}\label{metricS26}
ds^2=(1-\frac{\kappa M^4}{3}r^2)dt^2-(1+\frac{\kappa
M^4}{3}r^2-\frac{\kappa M^4}{2\lambda r^2})dr^2-r^2d\Omega^2.
\end{equation}

\noindent The metric (\ref{metricS26}) can be expressed by the
form of Newtonian potential

\begin{equation}
ds^2=(1+2\Phi)dt^2-\bigg[1-2\Phi\bigg(1-\frac{3}{2}(\frac{\delta}{r})^4\bigg)\bigg]dr^2-r^2d\Omega^2.
\end{equation}

\noindent It is obvious that the metric (\ref{metricS26}) is
linearized approximation of an AdS space-time.

\vspace{0.5cm}\noindent \textbf{4.2 Coupled Einstein and tachyon
field equations}\vspace{0.5cm}

In the above subsection, the linearized approximation of gravity
is used, which is available as long as the Newtonian potential is
much smaller than unit. In fact, for the spherical symmetric
metric (\ref{metric}), there is a general solution of the Einstein
equations with spherically symmetric energy-momentum tensor
$T^\mu_\nu$ which takes the form as \cite{li3,li4}

\begin{equation}\label{27}
A(r)^{-1}=1-\frac{\kappa}{r}\int^r_0T^0_0r^2dr
\end{equation}

\noindent and

\begin{equation}
B(r)=A(r)^{-1}\exp[\kappa\int^r_\infty (T^0_0-T^1_1)A(r)rdr].
\end{equation}

\noindent Substututjion of Eq.(\ref{t00}) on the Eq.(\ref{27})
yields

\begin{equation}\label{Aiterative}
A^{-1}=1-\frac{\kappa M^4}{r}\int^r_0(1+3\lambda
f^4)^{1/6}\exp(-\lambda
f^4)\sqrt{1+\frac{f'^2}{A}+\frac{2f^2}{r^2}}r^2dr,
\end{equation}

\noindent which is an integral equation for $A^{-1}$. Hence, we
want to use the method of successive approximation which provides
the solution as the limit of an infinite sequence of steps if
these steps were carried out exactly, simply and almost
identically each to the next so that the program is a relatively
easy task. For the small $\kappa M^4$, we take $A^{-1}=1$ as the
leading approximate value. Fortunately, in this case, we have an
analytical solution of $f(r)$, Eq.(\ref{f1}). Therefore, by direct
integrating Eq.(\ref{Aiterative}), we have

\begin{eqnarray}\label{sA2}
A^{-1}&=&1-\frac{\kappa
M^4}{r}\int^r_0\bigg[1+3\bigg(\frac{\delta}{r}\bigg)^4\bigg]^{2/3}\exp\bigg[-\bigg(\frac{\delta}{r}\bigg)^4\bigg]r^2dr\nonumber\\
&\simeq &1-\frac{\kappa M^4}{3}r^2+\frac{\kappa
M^4}{r}\bigg[\delta^3+g(\delta)\bigg]-\frac{\kappa
M^4}{\lambda}\frac{1}{r^2}
\end{eqnarray}

\noindent for $r>\delta$, where

\begin{equation}
g(\delta)=\int^{\delta}_0\bigg[1-\bigg[1+3\bigg(\frac{\delta}{r}\bigg)^4\bigg]^{2/3}
\exp\bigg[-\bigg(\frac{\delta}{r}\bigg)^4\bigg]\bigg]r^2dr
\end{equation}

\noindent is a monotonously increasing and positive (but smaller
than $\delta^3$) function. Analogously,

\begin{eqnarray}\label{sB2}
B&=&A^{-1}\exp\bigg[\kappa M^4\int^r_{\infty}\bigg(\frac{\delta}{
r}\bigg)^4\bigg[1+3\bigg(\frac{\delta}{r}\bigg)^4\bigg]^{-1/3}\exp\bigg[-\bigg(\frac{\delta}{
r}\bigg)^4\bigg]rdr\bigg]\nonumber\\
&\simeq &1-\frac{\kappa M^4}{3}r^2+\frac{\kappa
M^4}{r}\bigg[\delta^3+g(\delta)\bigg]-\frac{3\kappa
M^4}{2\lambda}\frac{1}{r^2}.
\end{eqnarray}

\noindent Eqs.(\ref{sA2}) and (\ref{sB2}) are the second
approximate value of the metric which are beyond those obtained by
linearized gravity approximation. It is easy to find that the
gravitational field of tachyon monopole is strongly repulsive and
the space-time becomes singular at a finite distance from the
monopole core. From aforementioned analysis, we find that metric
functions $A^{-1}(r)$ and $B(r)$ grow quadratically with the
distance off its core, therefore the metric represents one of
asymptotically AdS space-time. Furthermore, there is a relatively
small gravitational potential of repulsive nature, corresponding
to a mass $\sim M^4[\delta^3+g(\delta)]$ at the origin, but with
opposite sign.

\vspace{0.5cm}\noindent \textbf{5. Conclusions}\vspace{0.5cm}

In this paper, we have studied the property and gravitational
field of global monopole of tachyon matter in a four dimensional
approximately spherically-symmetric space-time. We give an
analytical solution of the tachyon field in the flat space-time
background and in particular, using the linearized approximation
of gravity and the method of successive approximation, we find the
approximate solution of the metric, which denotes an AdS
space-time with a negative mass at the origin. In contrast to the
ordinary monopole, the tachyon monopoles are very diffuse objects
with most of the energy distributed at large distances from the
monopole core, and their space-time is substantially distinct from
that of the ordinary monopole.

In previous papers \cite{li4}, Li and Hao have investigated the
ordinary global monopole in asymptotically dS/AdS space-time and
find that the mass of the monopole in the asymptotically dS
space-time could be positive if the cosmological constant is
greater than a critical value. This shows that the gravitational
field of the ordinary global monopole could be attractive or
repulsive depending on the value of the cosmological constant. We
note the cardinal difference between the gravitational effects of
tachyon and ordinary global monopole: for an ordinary global
monopole, its main effect, due to its mass that grows linearly
with the distance off its core, can be understood in terms of a
deficit solid angle \cite{barriola}; for a tachyon monopole, due
to its metric functions $A^{-1}(r)$ and $B(r)$ that grow
quadratically with the distance off its core, can be understood in
terms of asymptotically AdS space-time. From our analysis of the
coupled Einstein and tachyon field equations, we conclude that
there is a relatively small gravitational potential of repulsive
nature, corresponding to a negative mass at the origin. It is very
interesting to study the motion of a test particle and light in
the gravitational field of tachyon monopole. Although the test
particle motion is quite different from the case of the
Schwarzschild space-time with negative mass at the origin, the
behaviors of light remain unchanged except a energy (frequency)
shift. Some consequences of the existence of monopole with
negative masses have been considered long ago \cite{harari,
li1}.In the general relativity, Bondi \cite{bondi} showed that a
system moves with constant proper acceleration, asymptotically
approaching the speed of light. This runaway solution could be
seen as a drawback for a theory that allows particle with negative
mass. However, the tachyon monopoles are not isolated point
particles, but extended sources, even though the effective
negative mass reaches its asymptotic value very quickly outside
the core. One cannot expect a global monopole by itself, but
rather monopoles and anti-monopoles, which probably annihilated
themselves very efficiently, so that the motion of a system with a
monopole and a particle with equal positive mass will not runaway.

In the spirit of the string theory, the over all scale in the
tachyon potential is $M^4\propto \frac{1}{(2\pi)^3g_sl_s^4}$ and
$\lambda^{1/4} \propto \frac{1}{l_s}$, where $g_s$ and $l_s$ are
the coupling parameter and length scale of string, respectively.
The bound on $\lambda$ and $M$ by Eq.(\ref{21}) makes $g_s>>1$
which puts the filed theory description of string tachyons under
doubt. Moreover, the bound also makes the tachyon very light that
is an unacceptable situation from string theory point of
view\cite{Fairbairn}. Therefore, it is not suitable for tachyon
inspired from string theory to use linear approximation. However,
if we adopt the point of view that the tachyon potential,
Eq.(\ref{V1}) for example, comes purely from phenomenological
context just as those considered in Ref.\cite{Padmanabhan}, there
is no such problems.

Finally, it is worth noting that although we used the potential
(\ref{V1}) in our discussion, all the conclusions will be
preserved qualitatively if we use the potentials discussed by Sen
\cite{Sen2}, because these potentials have the similar quality as
the potential.

\vspace{0.8cm} \noindent \textbf{Acknowledgment}

This work is supported by National Natural Science Foundation of
China under Grant No. 10473007 and Shanghai Municipal Education
Commission (No.04DC28).

\end{document}